\documentclass[11pt,a4paper]{article}
\usepackage{amsmath}
\usepackage{amssymb}
\usepackage{amscd}
\usepackage[dvips]{graphicx}

\numberwithin{equation}{section}

\begin{document}

\newtheorem{thm}{Theorem}[section]
\newtheorem{cor}{Corollary}[section]
\newtheorem{lem}{Lemma}[section]
\newtheorem{prop}{Proposition}[section]
\newtheorem{defn}{Definition}[section]
\newtheorem{exas}{Examples}[section]
\newtheorem{exam}{Example}[section]
\newtheorem{counterexam}{Counterexample}[section]
\newtheorem{rem}{Remark}[section]
\newtheorem{ques}{Question}[section]
\newtheorem{conj}{Conjecture}[section]
\numberwithin{equation}{section}

\def\phi{\varphi}
\def\epsilon{\varepsilon}

\def\wt{\widetilde}
\def\fB{\mathfrak B}\def\fM{\mathfrak M}\def\fX{\mathfrak X}
 \def\cB{\mathcal B}\def\cM{\mathcal M}\def\cX{\mathcal X}
\def\mbe{\mathbf e}
\def\bu{\mathbf u}\def\bv{\mathbf v}\def\bx{\mathbf x} \def\by{\mathbf y} \def\bz{\mathbf z}
\def\om{\omega} \def\Om{\Omega}
\def\bbP{\mathbb P} \def\hw{h^{\rm w}} \def\hwi{{h^{\rm w}}}
\def\be{\begin{eqnarray}} \def\ee{\end{eqnarray}}
\def\beqq{\begin{eqnarray*}} \def\eeqq{\end{eqnarray*}}
\def\rd{{\rm d}} \def\Dwphi{{D^{\rm w}_\phi}}
\def\BX{\mathbf{X}}\def\Lam{\Lambda}\def\BY{\mathbf{Y}}
\def\BZ{\mathbf{Z}} \def\BN{\mathbf{N}}

\def\mwe{{D^{\rm w}_\phi}}
\def\DwPhi{{D^{\rm w}_\Phi}} \def\iw{i^{\rm w}_{\phi}}
\def\bE{\mathbb{E}}
\def\1{{\mathbf 1}} \def\fB{{\mathfrak B}}  \def\fM{{\mathfrak M}}
\def\diy{\displaystyle} \def\bbE{{\mathbb E}} \def\bu{\mathbf u}
\def\BC{{\mathbf C}} \def\lam{\lambda} \def\bbB{{\mathbb B}}
\def\bbR{{\mathbb R}}\def\bbS{{\mathbb S}}
 \def\bmu{{\mbox{\boldmath${\mu}$}}}
 \def\bPhi{{\mbox{\boldmath${\Phi}$}}}  \def\bPi{{\mbox{\boldmath{$\Pi$}}}}
  \def\btheta{{\mbox{\boldmath${\theta}$}}}
 \def\bbZ{{\mathbb Z}} \def\fF{\mathfrak F}\def\mbt{\mathbf t}\def\B1{\mathbf 1}
\def\hwphi{h^{\rm w}_{\phi}}
\def\BW{\mathbf{W}} \def\bw{\mathbf{w}}

\def\beal{\begin{array}{l}}
\def\beac{\begin{array}{c}}
\def\beacl{\begin{array}{cl}}
\def\ena{\end{array}}
\def\WBJ{\mathbf{J}^{\rm w}_{\phi}}
\def\BS{\mathbf{S}}
\def\BK{\mathbf{K}}
\def\BB{\mathbf{B}}
\def\wtD{{\widetilde D}}

\def\mwe{{D^{\rm w}_\phi}}
\def\DwPhi{{D^{\rm w}_\Phi}} \def\iw{i^{\rm w}_{\phi}}
\def\bE{\mathbb{E}}
\def\1{{\mathbf 1}} \def\fB{{\mathfrak B}}  \def\fM{{\mathfrak M}}
\def\diy{\displaystyle} \def\bbE{{\mathbb E}} \def\bu{\mathbf u}
\def\BC{{\mathbf C}} \def\lam{\lambda}
\def\bbB{{\mathbb B}} \def\bbM{{\mathbb M}}
\def\bbR{{\mathbb R}}\def\bbS{{\mathbb S}}
\def\blam{{\mbox{\boldmath${\lambda}$}}}
\def\bmu{{\mbox{\boldmath${\mu}$}}} \def\bta{{\mbox{\boldmath${\eta}$}}}
\def\bzeta{{\mbox{\boldmath${\zeta}$}}}
 \def\bPhi{{\mbox{\boldmath${\Phi}$}}}  \def\bPi{{\mbox{\boldmath{$\Pi$}}}}
 \def\bbZ{{\mathbb Z}} \def\fF{\mathfrak F}\def\mbt{\mathbf t}\def\B1{\mathbf 1}
\def\hwphi{h^{\rm w}_{\phi}}
\def\BT{{\mathbf T}} \def\BW{\mathbf{W}} \def\bw{\mathbf{w}}

\def\beal{\begin{array}{l}}
\def\beac{\begin{array}{c}}
\def\beacl{\begin{array}{cl}}
\def\ena{\end{array}}
\def\WBJ{\mathbf{J}^{\rm w}_{\phi}}
\def\BS{\mathbf{S}}
\def\BK{\mathbf{K}}
\def\tL{\mathbf{L}}
\def\BB{\mathbf{B}}
\def\vphi{{\varphi}}
\def\rw{{\rm w}}
\def\bZ{\mathbf Z}
\def\wtf{{\widetilde f}} \def\wtg{{\widetilde g}} \def\wtG{{\widetilde G}}
\def\vphi{\varphi}
\def\rT{{\rm T}}
\def\tA{{\tt A}} \def\tB{{\tt B}} \def\tC{{\tt C}} \def\tI{{\tt I}} \def\tJ{{\tt J}} \def\tK{{\tt K}}
\def\tL{{\tt L}} \def\tP{{\tt P}} \def\tQ{{\tt Q}} \def\tS{{\tt S}}
\def\beac{\begin{array}{c}} \def\beal{\begin{array}{l}} \def\beacl{\begin{array}{cl}} \def\ena{\end{array}}
\title{On double truncated (interval) WCRE and WCE}

\author{S. Yasaei Sekeh$^*$, G. R. Mohtashami Borzadran$^\dag$, A. H. Rezaei Roknabadi$^\ddag$}
\date{}
%\footnotetext{2010 {\em Mathematics Subject Classification:\;60A10, 60B05, 60C05}}
%\footnotetext{{\em Key words and phrases:} weighted cumulative entropy, double truncated (interval) weighted cumulative (residual) entropy, weight function\par}
\maketitle

\begin{abstract}
Measure of the weighted cumulative entropy about the predictability of failure time of a system have been introduced in \cite{M}. Referring properties of doubly truncated (interval) cumulative residual and past entropy, several bounds and assertions are proposed in weighted version.
\end{abstract}

{\small \textbf{2000 MSC.} 62N05, 62B10}\\
{\small\textbf{Keywords:} weighted cumulative entropy, double truncated (interval) weighted cumulative (residual) entropy, weight function.}
\footnote{$^*$Department\;of\;Statistics,\;Federal\;University\;of\;S$\tilde{\rm a}$o\;Carlos\;(UFSCar),\;S$\tilde{\rm a}$o\;Carlos,\;Brazil.\;E-mail: sa$_{-}$yasaei@yahoo.com}
\footnote{$^\dag$Department\;of\;Statistics,\;Ferdowsi\;University\;of\;Mashhad,\;Mashhad,\;Iran.\;E-mail: gmb1334@yahoo.com;\\ grmohtashami@um.ac.ir}\\
\footnote{$^\ddag$Department\;of\;Statistics,\;Ferdowsi\;University\;of\;Mashhad,\;Mashhad,\;Iran.\;E-mail: rezaei494@gmail.com}
%\mbox{\quad}
\def\fB{\mathfrak B}\def\fM{\mathfrak M}\def\fX{\mathfrak X}
 \def\cB{\mathcal B}\def\cM{\mathcal M}\def\cX{\mathcal X}
\def\bu{\mathbf u}\def\bv{\mathbf v}\def\bx{\mathbf x}\def\by{\mathbf y}
\def\om{\omega} \def\Om{\Omega}
\def\bbP{\mathbb P} \def\hw{h^{\rm w}} \def\hwphi{{h^{\rm w}_\phi}} \def\bbR{\mathbb R}
\def\beq{\begin{eqnarray}} \def\eeq{\end{eqnarray}}
\def\beqq{\begin{eqnarray*}} \def\eeqq{\end{eqnarray*}}
\def\rd{{\rm d}} \def\Dwphi{{D^{\rm w}_\phi}}
\def\BX{\mathbf{X}}
\def\mwe{{D^{\rm w}_\phi}}
\def\DwPhi{{D^{\rm w}_\Phi}} \def\iw{i^{\rm w}_{\phi}}
\def\bE{\mathbb{E}}
\def\1{{\mathbf 1}} \def\fB{{\mathfrak B}}  \def\fM{{\mathfrak M}}
\def\diy{\displaystyle} \def\bbE{{\mathbb E}}
\def\bF{\overline{F}}
\def\ew{{\mathcal{E}^{\rm w}_{\phi}}}
\def\cew{{\overline{\mathcal{E}}^{\rm w}_{\phi}}}
\let\phi\varphi
\section{Introduction. Interval weighted cumulative entropies}

Let $x\in \bbR^+\mapsto \phi(x)\geq 0$ be a given measurable function.
The weighted cumulative residual entropy (WCRE) $\ew(X)$ and the weighted cumulative entropy (WCE) $\cew(X)$ of a RV $X$ with a cumulative distribution function (CDF) $F$ and survival function (SF) $\bF$ are defined by
\beq\label{eq:WCRE}
\ew (X)=\ew (F) =-\int_{\bbR^+}\phi (x )\bF(x)\log\,\bF(x) \rd x,\;\;\; \hbox{and}\eeq
\beq\label{eq:WCE}
\cew (X)=\cew (F) =-\int_{\bbR^+}\phi (x )F(x) \log\,F(x)\rd x, \eeq
respectively. Assume that all integrals are absolutely convergent with the standard agreement $0\log 0=0\log \infty=0$. Cf. \cite{M}, \cite{CL} and \cite{RCVW}. Further for more details and motivations see \cite{SY}, \cite{Ys}.

For given pair of fixed values $(t_1,t_2)\in \bbR^+\times \bbR^+$ the CDF $F(x;t_1,t_2)$ and SF $\bF(x;t_1,t_2)$ of a RV $X|t_1<X<t_2$ take the forms
\beq\label{eq:Int.03} F(x;t_1,t_2)=\frac{F(x)}{F(t_2)-F(t_1)}\quad\;\;\ \hbox{and}\;\; \quad \bF(x;t_1,t_2)=\frac{\bF(x)}{\bF(t_1)-\bF(t_2)}.\eeq

 We propose the following definition which we call the double truncated (interval) weighted cumulative residual entropy (IWCRE) $I\ew(t_1,t_2)$ and the double truncated (interval) weighted cumulative entropy (IWCE) $I\cew(t_1,t_2)$ of a RV $X|t_1<X<t_2$:
\begin{defn}\label{defnn:1}
Let $(t_1,t_2)$ be a pair of fixed values in $\bbR^+\times \bbR^+$. Using (\ref{eq:Int.03}) define IWCRE of a RV $X|t_1<X<t_2$ with SF $\bF$ and WF $\phi$ by:
\beq \label{def.IWCRE} \begin{array}{ccl} I\ew(t_1,t_2)&=&\diy -\int_{t_1}^{t_2} \phi(x) \bF(x;t_1,t_2) \log\; \bF(x;t_1,t_2) \rd x\\
&=&-\diy\int_{t_1}^{t_2} \phi(x) \frac{\bF(x)}{\bF(t_1)-\bF(t_2)} \log\; \frac{\bF(x)}{\bF(t_1)-\bF(t_2)} \rd x,\end{array}\eeq
and the IWCE of a RV $X|t_1<X<t_2$ with CDF $F$ is defined by
\beq \label{def. IWCE} \begin{array}{ccl} I\cew(t_1,t_2)&=&-\diy \int_{t_1}^{t_2} \phi(x) F(x;t_1,t_2) \log\; F(x;t_1,t_2) \rd x\\
&=&-\diy\int_{t_1}^{t_2} \phi(x) \frac{F(x)}{F(t_2)-F(t_1)}\log\; \frac{F(x)}{F(t_2)-F(t_1)} \rd x.\end{array}\eeq
\end{defn}
In particular $\phi(x)\equiv1$ the (\ref{def.IWCRE}) and (\ref{def. IWCE}) yield the standard Interval cumulative residual entropy and the interval cumulative entropy, respectively. Cf. \cite{KRM}, \cite{NR}, \cite{MY} and \cite{SSM}.\\

Passing to the limits $t_1\rightarrow 0$ and $t_2\rightarrow \infty$, the IWCRE (\ref{def.IWCRE}) and IWCE (\ref{def. IWCE}) intend the WCRE (\ref{eq:WCRE}) and the WCE (\ref{eq:WCE}), that is $I\ew(0,\infty)=\ew(X)$ and $I\cew(0,\infty)=\cew(X)$.
\begin{rem}
{\rm (a)} Owing to Definition \ref{defnn:1}, assume exponential RV $X$, $X\sim Exp(\lambda),\;\lambda\geq0$. In particular for given real constants $a_0,\dots,a_n$ where $\varphi(x)=\diy\sum_{i=0}^n a_i\; x^i\geq 0$. Set
\beq\label{def:gamma} \gamma(b,z)=\diy\int_0^z t^{b-1}\;e^{-t}\;\rd t.\eeq
\def\lam{\lambda}
Following some straightforward computations one obtains
\beq\label{Eq;1.7} \begin{array}{l} I\ew(t_1,t_2)= \diy \Big(e^{-t_1/\lam}-e^{-t_2/\lam}\Big)^{-1}\diy\sum_{i=0}^n a_i\;\lam^{i+1}\Big\{\gamma(i+2,t_2/\lam)-\gamma(i+2,t_1/\lam)\Big\}\\
\quad+\diy\Big(e^{-t_1/\lam}-e^{-t_2/\lam}\Big)^{-1}\log \Big(e^{-t_1/\lam}-e^{-t_2/\lam}\big)\diy\sum_{i=0}^n a_i\;\lam^{i}\Big\{\gamma(i+1,t_2/\lam)-\gamma(i+1,t_1/\lam)\Big\}.\end{array}\eeq
{\rm (b)} More generally, let $\mu\in\bbR$, $\sigma>0$, $\xi\in\bbR^+$ such that $\mu-\sigma/\xi\geq 0$ be location, scale and shape parameters respectively. Suppose that RV $X$ has $GEV(\mu,\sigma,\xi)$ distribution, with CDF
\beq F_{GEV}(x)=e^{-y(x)},\quad \hbox{where}\quad y(x)=\big(1+\big(\diy\frac{x-\mu}{\sigma}\big)\xi\big)^{-1/\xi}.\eeq
Moreover, set
\beq\label{def:Pi} \Pi_c(a,b)=\diy\int_a^b y(t)^{c-1}e^{-y(t)}\rd t,\quad a,b>0,\;\; c\in\bbR.\eeq
If we assume $\varphi(x)=\diy\sum_{i=0}^n b_i\; y(x)^i$, for $b_i\in \bbR,\;i=0\dots n$ such that $\varphi(x)\geq 0$ with obvious motivations, the following expression is derived:
\beq \begin{array}{l} I\cew(t_1,t_2)=\diy\Big(e^{-y(t_2)}-e^{-y(t_1)}\Big)^{-1}\sum_{i=0}^n b_i\Pi_{i+2}(t_1,t_2)\\
\quad+\diy \Big(e^{-y(t_2)}-e^{-y(t_1)}\Big)^{-1}\log \Big(e^{-y(t_2)}-e^{-y(t_1)}\Big)\sum_{i=0}^n b_i \Pi_{i+1}(t_1,t_2).\end{array}\eeq
\end{rem}

From now on for given WF $\phi$ we will use the notation $\psi(x)=\diy \int_0^x \phi(s)\rd s$.
\vskip .5 truecm
The following  Lemma is straightforward.
\begin{lem}  For given a pari $(t_1,t_2)$ and WF $\phi$ applying integrate by parts in Eqn (\ref{def.IWCRE}) and (\ref{def. IWCE}) it can be written equivalent forms for IWCRE and IWCE:
\beq \label{eq:Int.05} \begin{array}{ccl}
I\ew(t_1,t_2)&=& \diy\frac{1}{\bF(t_2)-\bF(t_1)}\int_{t_1}^{t_2} \phi(x) \bF(x)\log\;\bF(x) \rd x+{\bar{\delta}}^{\rm w}_{\phi}(t_1,t_2)\log \{ \bF(t_1)-\bF(t_2)\}\\
&=&\diy \frac{1}{\bF(t_2)-\bF(t_1)}\int_{t_1}^{t_2} \phi(x) \bF(x)\log\;\bF(x) \rd x\\
&&\quad+\diy\left\{\frac{\psi(t_2)\bF(t_2)-\psi(t_1)\bF(t_1)}{\bF(t_1)-\bF(t_2)}+\bbE\big[\psi(X)|t_1<X<t_2\big]\right\}\log\{\bF(t_1)-\bF(t_2)\},
\end{array}\eeq
and in similar way:
\beq \label{eq:Int.06} \begin{array}{ccl}
I\cew(t_1,t_2)&=& \diy\frac{1}{F(t_1)-F(t_2)}\int_{t_1}^{t_2} \phi(x) F(x)\log\;F(x) \rd x+\delta^{\rm w}_{\phi}(t_1,t_2)\log \{ F(t_2)-F(t_1)\}\\
&=&\diy \frac{1}{F(t_1)-F(t_2)}\int_{t_1}^{t_2} \phi(x) F(x)\log\;F(x) \rd x\\
&&\quad+\diy\left\{\frac{\psi(t_2)F(t_2)-\psi(t_1)F(t_1)}{F(t_2)-F(t_1)}-\bbE\big[\psi(X)|t_1<X<t_2\big]\right\}\log\{F(t_2)-F(t_1)\}.
\end{array}\eeq
Here
\beq\label{eq.deltas} {\bar{\delta}}^{\rm w}_{\phi}(t_1,t_2)=\int_{t_1}^{t_2} \phi(x)\frac{\bF(x)}{\bF(t_1)-\bF(t_2)}\rd x \quad ,\quad \delta^{\rm w}_{\phi}(t_1,t_2)=\int_{t_1}^{t_2} \phi(x)\frac{F(x)}{F(t_2)-F(t_1)} \rd x.\eeq
\end{lem}

Setting $\phi'(x)$ the derivative function of WF $\phi(x)$ with respect to $x$, $\phi'(x)=\frac{\partial}{\partial x}\phi(x)$ and following some standard calculations, we can write:
\beq \label{eq:Int.04} \begin{array}{c}I\cew(t_1,t_2)=\diy \phi(t_1)\overline{\mathcal{E}}_X(t_1,t_2)+\int_{t_1}^{t_2} \phi'(x) \overline{\mathcal{B}}_X(x,t_2)\rd x, \\
I\cew(t_1,t_2)=\diy-\phi(t_2)\overline{\mathcal{E}}_X(t_1,t_2)+\int_{t_1}^{t_2} \phi'(y) \overline{\mathcal{B}}_X(t_1,y)\rd y,\end{array}\eeq
here $\overline{\mathcal{E}}_X(t_1,t_2)$ represents the interval cumulative past entropy, denoted by $ICPE(X;t_1,t_2)$, in \cite{KRM}. Moreover,
\beq \begin{array}{cl} \overline{\mathcal{B}}_X(x,t_2)=-\diy\int_x^{t_2} \frac{F(y)}{F(t_2)-F(t_1)}\log \frac{F(y)}{F(t_2)-F(t_1)}\rd y,\\
\overline{\mathcal{B}}_X(t_1,y)=-\diy\int_{t_1}^y \frac{F(x)}{F(t_2)-F(t_1)}\log \frac{F(x)}{F(t_2)-F(t_1)}\rd x. \end{array}\eeq
In (\ref{eq:Int.04}), substitute $\mathcal{E}_X(t_1,t_2)$ (denoted by $ICRE(X;t_1,t_2)$, cf. \cite{KRM}) in  $\overline{\mathcal{E}}_X(t_1,t_2)$, the analogue assertion for $I\ew(t_1,t_2)$ holds.

%Furthermore, take the derivative from both sides of (\ref{eq:Int.04}) with respect to $t_1$ and $t_2$. One can get a representation for the $I\cew(t_1,t_2)$:
%\beq\label{eq.derivative} \frac{\partial}{\partial t_1} I\cew(t_1,t_2)=\phi(t_1)\;\frac{\partial}{\partial t_1} \overline{\mathcal{E}}_X(t_1,t_2) \quad ,\quad \frac{\partial}{\partial t_2} I\cew(t_1,t_2)=-\phi(t_2)\;\frac{\partial}{\partial t_2} \overline{\mathcal{E}}_X(t_1,t_2).\eeq

%Again, straightforward calculations gives similar results for $I\ew(t_1,t_2)$.

%\begin{rem}
%According to (\ref{eq.derivative}), the IWCE is increasing (decreasing) in $t_1$ for given WF $\phi$ if and only if the cumulative past entropy, $\mathcal{CE}_X(t_1,t_2)$ is increasing (decreasing) in $t_1$ whereas IWCE is increasing (decreasing) in $t_2$ if and only if $\mathcal{CE}_X(t_1,t_2)$ is decreasing (decreasing) in $t_2$. This property also holds true for IWCRE.
%\end{rem}
\def\lam{\lambda}
\vskip .5 truecm
\begin{exam}
{\rm Let $X$ be a RV from exponential distribution with mean $\diy\frac{1}{\lam}$, $\lambda>0$. According to the example in the end of \cite{KRM}:
\beq\label{ICRE.exp}I\mathcal{E}(t_1,t_2)=\diy\frac{1}{\lam}+\frac{1}{\lam}\log \big(1-e^{\lam(t_1-t_2)}\big)+\frac{(t_2-t_1)e^{\lam t_1}}{e^{\lam t_1}-e^{\lam t_2}},\;\; t_2>t_1\geq 0.\eeq
We observe that for fixed value $t_2\in (0,\infty)$, (\ref{ICRE.exp}) is decreasing in $t_1\in (0,\infty)$. Now, assume the WF $\phi(x)=e^{\alpha x}$, $\alpha < \lam$, applying (\ref{def.IWCRE}) yields the following expression:
\beq \label{IWCRE.exp}\begin{array}{l}
I\ew (t_1,t_2)=\diy\frac{1}{(\lam-\alpha)(e^{-\lam t_2}-e^{-\lam t_1})}. \bigg\{ \lam \big(t_2 e^{(\alpha-\lam)t_2}-t_1e^{(\alpha-\lam)t_1}\big)\\
\quad+\diy\frac{\lam}{(\alpha-\lam)}.\big(e^{(\alpha-\lam)t_2}-e^{(\alpha-\lam)t_1}\big)+\big(e^{(\alpha-\lam)t_2}-e^{(\alpha-\lam)t_1}\big).\log \big(e^{-\lam t_1}-e^{-\lam t_2}\big)\bigg\}.\end{array}\eeq
Note that when $\alpha \rightarrow 0$ then $I\ew (t_1,t_2)\rightarrow I\mathcal{E}(t_1,t_2)$. Applying mathematical software such as Maple, one can easily check that for given all $\lambda$, $\alpha$, (\ref{IWCRE.exp}) is not monotonic decreasing in $t_1$. This means, if the monotonicity property for ICRE is fulfilled then there is no guarantee IWCRE is monotonic as well. }
\end{exam}

\section{Bounds for the IWCE and IWCRE}

$\quad$In this section , we give several bounds for the IWCRE and IWCE by using assertions established in Section 1.
Let us start with an alternative representation for the IWCRE and IWCE. In fact it follows the same line as (\ref{eq:Int.05}) and (\ref{eq:Int.06}) but is more elementary.\\

Let $X$ be a non-negative RV, moreover consider a pair $(t_1,t_2)\in \bbR^+\times \bbR^+$. Set
\beqq \gamma(t_1,t_2)=-\diy\int_{t_1}^{t_2}\varphi(x)\; F(x)\log F(x)\;\rd x,\\
\vartheta_i(t_1,t_2)=\diy\frac{F(t_i)}{F(t_2)-F(t_1)},\quad i=1,2.\eeqq
therefore, we can write
\beq \label{vartheta1}\begin{array}{ccl} I\cew (t_1,t_2)=&-&\diy \int_{t_1}^{t_2} \phi(x) \frac{F(x)}{F(t_2)-F(t_1)} \log\; \vartheta_1(x,t_2) \rd x\\
&-&\diy \int_{t_1}^{t_2} \phi(x) \frac{F(x)}{F(t_2)-F(t_1)} \log\;\frac{F(t_2)-F(x)}{F(t_2)-F(t_1)} \rd x,\end{array}\eeq
in addition,
\beq\label{vartheta2} \begin{array}{ccl} I\cew (t_1,t_2)=&-&\diy \int_{t_1}^{t_2} \phi(x) \frac{F(x)}{F(t_2)-F(t_1)} \log\; \vartheta_2(t_1,x) \rd x\\
&-&\diy \int_{t_1}^{t_2} \phi(x) \frac{F(x)}{F(t_2)-F(t_1)} \log\;\frac{F(x)-F(t_1)}{F(t_2)-F(t_1)} \rd x.\end{array}\eeq
For given pair $(t_1,t_2)$ define functions $\bar{\gamma}_1$ and $\bar{\gamma}_2$ in terms of $\bF(x)$ in a similar fashion, then analogue formulas take place for IWCRE as well.\\
Now we are in the position to establish Theorem \ref{thm2.1} below. Recalling (\ref{eq.deltas}), (\ref{vartheta1}), (\ref{vartheta2}) and using the inequality $\log (1-s)\geq s\big/(s-1),\; 0<s<1$ we provide lower bounds for the IWCE, omitting the proof.
\vskip .5 truecm
\begin{thm}\label{thm2.1}
Let $X$ be a non-negarive RV, with CDF $F$. Then given WF $x\in \bbR^+\mapsto\phi (x )\geq 0$ obeys
\beq\label{Eq:thm2.1}\begin{array}{l} I\cew(t_1,t_2)\geq\\
 \quad \diy\big(F(t_2)-F(t_1)\big)^{-1}\Big[\gamma(t_1,t_2)+F(t_2)\big(\psi(t_2)-\psi(t_1)\big)\Big]+\delta^{\rm w}_{\phi}(t_1,t_2)\Big(1+\log F(t_1)\Big).\end{array}\eeq
%\begin{itemize}
%\item[] (i) Assume that $\gamma_1(t_1,t_2)$ is decreasing in $t_1$, then
%\beqq I\cew(t_1,t_2)\geq - \delta^{\rm w}_{\phi}(t_1,t_2) \log\gamma_1(t_1,t_2),\eeqq
%\item[] (ii) Assume that $\gamma_2(t_1,t_2)$ is increasing in $t_2$, then
%\beqq I\cew(t_1,t_2)\geq - \delta^{\rm w}_{\phi}(t_1,t_2) \log\gamma_2(t_1,t_2).\eeqq
%\end{itemize}
It is worth noting that in a similar manner by owing to the definition of ${\bar{\delta}}^{\rm w}_{\phi}(t_1,t_2)$ in (\ref{eq.deltas}), if we swap $\gamma$ and $\overline{\gamma}$, also $F$ and $\bF$ in \ref{Eq:thm2.1} we get analogue lower bounds for $I\ew(t_1,t_2)$, where
\beqq \overline{\gamma}(t_1,t_2)=-\diy\int_{t_1}^{t_2}\varphi(x)\; \bF(x)\log \bF(x)\;\rd x.\eeqq
\vskip .5 truecm
\end{thm}

An immediate application of Theorem \ref{thm2.1} follows.
\begin{prop}
Consider function $g(\varepsilon)$ in a form as
\beqq g(\varepsilon)=\big(1+\big(\diy\frac{\varepsilon-\mu}{\sigma}\big)\xi\big)^{-1/\xi},\quad \sigma>0,\;\mu\in\bbR,\;\xi\in\bbR^+,\;\mu-\sigma/\xi\geq 0.\eeqq
Then for constant $0\leq x<y\leq \infty$ and $\theta_i$, $i=0\dots n$ the inequality
\beq\Big(g(x)-1+\log\big(e^{-g(y)}-e^{-g(x)}\big)\Big)\sum_{i=0}^n \theta_i\; \Pi_{i+1}(x,y)\geq e^{-g(y)} \sum_{i=0}^n \theta_i \diy\int_x^y g(s)^i\;\rd s.\eeq
holds true. Here $\Pi$ stands as before in (\ref{def:Pi}):
\beqq\Pi_c(a,b)=\diy\int_a^b y(t)^{c-1}e^{-y(t)}\rd t,\quad a,b>0,\;\; c\in\bbR.\eeqq
\end{prop}

\begin{thm}\label{thm.2.1}
Suppose that $X$ is a RV with CDF $F$ and finite $I\cew(t_1,t_2)$. Given WF $\phi$, set
\beqq\eta(X)=\diy\frac{1}{F(x)}\int_0^x \phi(y)F(y)\rd y\eeqq
Then
\beqq I\cew(t_1,t_2)\leq \bbE\left[\eta(X)|t_1\leq X \leq t_2\right].\eeqq
\end{thm}
\vskip .5 truecm
{\bf Proof.}\;First we begin from the expression $\eta(X)$:
\beqq \begin{array}{c}
\bbE\left[\eta(X)|t_1\leq X \leq t_2\right]=\diy\int_{t_1}^{t_2} \left(\int_0^x \phi(y) \frac{F(y)}{F(x)}\rd y\right)\frac{f(x)}{F(t_2)-F(t_1)}\rd x\\
\quad=\diy\int_{0}^{t_1}\left(\int_{t_1}^{t_2}\frac{f(x)}{F(x)}\rd x\right)\phi(y)\frac{F(y)}{F(t_2)-F(t_1)}\rd y + \int_{t_1}^{t_2}\left(\int_{y}^{t_2}\frac{f(x)}{F(x)}\rd x\right)\phi(y)\frac{F(y)}{F(t_2)-F(t_1)}\rd y.\end{array}\eeqq
Further using the relation $\diy\int_a^b \frac{f(x)}{F(x)}\rd x=\log F(b)-\log F(a)$ leads
\beq\label{eq.2.1} \begin{array}{l}
\bbE\Big[\eta(X)|t_1\leq X \leq t_2\Big]=\diy\int_0^{t_1}\big[\log F(t_2)-\log F(t_1)\big]\phi(y)\frac{F(y)}{F(t_2)-F(t_1)}\rd y\\
\qquad+\diy\int_{t_1}^{t_2} \big[\log F(t_2)-\log F(y)\big]\phi(y)\frac{F(y)}{F(t_2)-F(t_1)}\rd y\\
\qquad\geq\diy \int_{t_1}^{t_2} \big[\log \{F(t_2)-F(t_1)\}-\log F(y)\big]\phi(y)\frac{F(y)}{F(t_2)-F(t_1)}\rd y.\end{array}\eeq
In the last line of (\ref{eq.2.1}) the inequality holds from $\log F(t_2)-\log F(t_1) \geq 0$. For given $t_1<t_2\in\bbR^+$ we also know  $\log F(t_2)\geq \log \big[F(t_2)-F(t_1)\big]$. This completes the proof. $\quad$ $\blacksquare$
\vskip .5 truecm
Remarkably observe that, IWCRE possesses the similar property in Theorem \ref{thm.2.1}, hence we can write:
\beqq I\ew(t_1,t_2)\leq \bbE\left[\bar{\eta}(X)|t_1\leq X \leq t_2\right],\eeqq
where $\bar{\eta}(x)=\diy \frac{1}{\bF(x)}\int_x^\infty \phi(y) \bF(y)\rd y$.\\

The next theorem extends the result of Theorem 8 from \cite{KRM}. Here we set
\beqq IH(X;t_1,t_2)=-\int_{t_1}^{t_2}  \frac{f(x)}{F(t_2)-F(t_1)} \log \frac{f(x)}{F(t_2)-F(t_1)} \rd x,\eeqq
Note that $IH(X;t_1,t_2)$ is an extension of Shannon entropy based on a doubly truncated (interval) RV, see \cite {SSM}.
\vskip .5 truecm
\begin{thm}\label{thm2.3.}
Let $X$ be a non-negative continuous RV with PDF and CDF respectively $f(x)$ and $F(x)$, then for give WF $\phi(x)$,
\beqq I\cew(t_1,t_2)\geq \alpha(t_1,t_2). \exp\{IH(X;t_1,t_2)\}.\eeqq
Here
\beqq \alpha(t_1,t_2)=\exp \Big\{\int_{\beta_1}^{\beta_2} \log \big[u\; \phi(F^{-1}\{u F(t_2)-u F(t_1)\})|\log u| \big]\rd u \Big\},\eeqq
where for $i=1,2$, $\beta_i=\diy \frac{F(t_i)}{F(t_2)-F(t_1)}$.
\end{thm}
\vskip .5 truecm
{\bf Proof.}\; The proof follows directly from the Log-Sum inequality while implies
\beqq \begin{array}{l}
\diy\int_{t_1}^{t_2} \frac{f(x)}{F(t_2)-F(t_1)} \log \frac{f(x)}{F(t_2)-F(t_1)}\rd x\\
\quad-\diy\int_{t_1}^{t_2}\frac{f(x)}{F(t_2)-F(t_1)}\log \big[\phi(x)\frac{F(x)}{F(t_2)-F(t_1)}|\log \frac{F(x)}{F(t_2)-F(t_1)}|\big]\rd x \\
\qquad \geq -\diy\log \int_{t_1}^{t_2}\phi(x)\frac{F(x)}{F(t_2)-F(t_1)}|\log \frac{F(x)}{F(t_2)-F(t_1)}|\rd x\\
\qquad =\diy\log \frac{1}{I\cew(t_1,t_2)}.\qquad\qquad \qquad\hspace{4cm} \blacksquare\end{array}\eeqq
\vskip .5 truecm
\begin{rem}
The similar arguments for IWCRE is achieved. In other words, owing to the definition of $IH(X;t_1,t_2)$ we have
\beqq I\ew(t_1,t_2)\geq \bar{\alpha}(t_1,t_2). \exp\Big\{IH(X;t_1,t_2)\Big\}.\eeqq
Here
\beqq \bar{\alpha}(t_1,t_2)=\exp \Big\{\int_{\kappa_1}^{\kappa_2} \log \big[u\; \phi(\bF^{-1}\{u \bF(t_1)-u \bF(t_2)\})|\log u| \big]\rd u \Big\},\eeqq
where for $i=1,2$, $\kappa_i=\diy \frac{\bF(t_i)}{\bF(t_1)-\bF(t_2)}$.
\end{rem}
\vskip .5 truecm

In Theorem \ref{thm:bound.monotonic} below (cf. Theorem 2.3, \cite{KRM}), let $\bar{\lambda}(x)=\diy\frac{f(x)}{F(x)}$ be reversed failure rate function and $h_2(t_1,t_2)$ denotes the generalized failure rate (GFR) by virtue of the doubly truncated RV, defined in \cite{NR}. Assume also $\phi(x)$ be a positive WF on an open domain with $\psi(x)=\diy\int_0^x \phi(s)\rd s$ and set $\mathcal{M}(t_1,t_2)=\bbE\Big[\psi(t_2)-\psi(X)|t_1\leq X \leq t_2\Big]$. Then the next theorem is provided:
\vskip .5 truecm
\begin{thm} \label{thm:bound.monotonic}
The IWCE is an increasing function in $t_2$ iff for all given $(t_1,t_2)\in \bbR^+\times \bbR^+$, $t_1<t_2$:
\beq\label{eq:bound.monottonic} \begin{array}{l}I\cew(t_1,t_2)\\
\quad\leq \diy \mathcal{M}(t_1,t_2)+(\psi(t_2)- \psi(t_1))\frac{F(t_1)}{F(t_2)-F(t_1)}-\phi(t_2)[\bar{\lambda}(t_2)]^{-1} \log \frac{F(t_2)}{F(t_2)-F(t_1)}.\end{array}\eeq
\end{thm}
\vskip .5 truecm
{\bf Proof.}\; According to the form (\ref{eq:Int.06}), differentiating IWCE with respect to $t_2$ yields
\beq\label{eq:sec2.1} \begin{array}{l}\diy\frac{\partial}{\partial t_2} I\cew(t_1,t_2)=\diy \frac{f(t_2)}{[F(t_2)-F(t_1)]^2}\int_{t_1}^{t_2} \phi(x) F(x) \log F(x) \rd x -\frac{\phi(t_2)F(t_2)\log F(t_2)}{F(t_2)-F(t_1)}\\
\qquad+ \diy \frac{f(t_2)}{F(t_2)-F(t_1)}\bigg[\mathcal{M}(t_1,t_2)+(\psi(t_2)-\psi(t_1))\frac{F(t_1)}{F(t_2)-F(t_1)}\bigg]\\
\qquad+\diy \bigg(\frac{\partial}{\partial t_2}\mathcal{M}(t_1,t_2)+\frac{\phi(t_2) F(t_1)}{F(t_2)-F(t_1)}-\frac{f(t_2)F(t_1) (\psi(t_2)-\psi(t_1))}{[F(t_2)-F(t_1)]^2}\bigg) \log \{F(t_2)-F(t_1)\}.\end{array}\eeq
Furthermore differentiating the $\mathcal{M}(t_1,t_2)$ with respect to $t_2$ implies
\beq\label{eq.2.2} \frac{\partial}{\partial t_2}\mathcal{M}(t_1,t_2)=\phi(t_2)-\mathcal{M}(t_1,t_2)h_2(t_1,t_2).\eeq
After that substitute (\ref{eq.2.2}) in (\ref{eq:sec2.1}), we have
\beqq \begin{array}{l}\diy \frac{\partial}{\partial t_2} I\cew(t_1,t_2)\\
\quad=\diy h_2(t_1,t_2).\bigg[\mathcal{M}(t_1,t_2)-I\cew(t_1,t_2)+\diy(\psi(t_2)- \psi(t_1))\frac{F(t_1)}{F(t_2)-F(t_1)}\\
\qquad\qquad-\diy\phi(t_2)[\bar{\lambda}(t_2)]^{-1} \log \frac{F(t_2)}{F(t_2)-F(t_1)}\bigg].\end{array}\eeqq
The inequality (\ref{eq:bound.monottonic}) then follows. $\quad$ $\blacksquare$
\vskip .5 truecm
\begin{thm}\label{thm.sec2.2}
{\rm{(Cf. \cite{KRM} Theorem 2.10)}} Suppose $X$ and $Y$ are two non-negative, iid RVs with SF $\bF$. Then for given WF $\phi$ , consequently $\psi$ and $(t_1,t_2)\in \bbR^+\times\bbR^+$, $t_1<t_2$:
\beq \label{eq:Sec2.3} \begin{array}{l}
\diy \bbE\Big(|\psi(X)-\psi(Y)|| t_1\leq X \leq t_2,t_1\leq Y \leq t_2 \Big)\\
\\
\quad \leq \diy \frac{2 I\ew(t_1,t_2)}{\bF(t_1)-\bF(t_2)} - \frac{\log[\bF(t_1)-\bF(t_2)]}{\bF(t_1)-\bF(t_2)}\bigg(\overline{\mathcal{M}}(t_1,t_2)+(\psi(t_2)-\psi(t_1))\frac{\bF(t_2)}{\bF(t_1)-\bF(t_2)}\bigg).
\end{array}\eeq
Here
$$\overline{\mathcal{M}}(t_1,t_2)=\bbE\Big[\psi(X)-\psi(t_1)|t_1\leq X\leq t_2\Big].$$
\end{thm}
\vskip .5 truecm
{\bf Proof.}\;  Following the similar arguments in Theorem 2.10, \cite{KRM}, for two iid RVs $X$ and $Y$ we have
\beq\label{eq2.4} \begin{array}{l}
\diy 2\;\frac{\bF(u)}{\bF(t_1)-\bF(t_2)}-2\;\bigg(\frac{\bF(u)}{\bF(t_1)-\bF(t_2)}\bigg)^2\\
\\
\qquad\quad=\diy P\{\max(\phi(X),\phi(Y))>u|t_1\leq X \leq t_2,t_1\leq Y \leq t_2\}\\
\\
\qquad\qquad- \diy P\{\min(\phi(X),\phi(Y))>u|t_1\leq X \leq t_2,t_1\leq Y \leq t_2\}.\end{array}\eeq
By multiplying the both sides of (\ref{eq2.4}) in $\phi(u)$ and then integrating from $t_1$ to $t_2$, we obtain
\beqq \begin{array}{l} \diy \frac{2}{[\bF(t_1)-\bF(t_2)]^2}\int_{t_1}^{t_2} \phi(u) \bF(u)\left[\bF(t_1)-\bF(t_2)-\bF(u)\right] \rd u\\
\\
\qquad =\diy \bbE\Big(|\psi(X)-\psi(Y)|| t_1\leq X \leq t_2,t_1\leq Y \leq t_2 \Big).\end{array}\eeqq
At this stage we apply the non-decreasing property for $\psi$ in $x$ and deduce that for all $x\in(0,1)$ and $b\in(0,1)$, $x(b-x)\leq x|\log x|$. This leads to
\beq\label{eq2.5} \begin{array}{l}\diy\bbE\Big(|\psi(X)-\psi(Y)|| t_1\leq X \leq t_2,t_1\leq Y \leq t_2 \Big)\\
 \qquad\leq \diy \frac{2}{[\bF(t_1)-\bF(t_2)]^2}\int_{t_1}^{t_2} \phi(u) \bF(u)|\log \bF(u)| \rd u.\end{array}\eeq
Combining (\ref{eq2.5}) and (\ref{eq:Int.05}) the assertion (\ref{eq:Sec2.3}) clarifies. $\quad$ $\blacksquare$
\vskip .5 truecm
\begin{rem}
It can be observed explicitly that the LHS of inequality (\ref{eq:Sec2.3}) in Theorem \ref{thm.sec2.2} is bigger and equal than:
\beqq \bbE\Big(|\psi(X)-\bbE(\psi(X))|| t_1\leq X \leq t_2\Big).\eeqq
Moreover, similar inequalities as (\ref{eq:Sec2.3}) for IWCE can be hold:
\beq \begin{array}{l}
\diy \bbE\Big(|\psi(X)-\psi(Y)|| t_1\leq X \leq t_2,t_1\leq Y \leq t_2 \Big)\\
\\
\quad \leq \diy \frac{2 I\cew(t_1,t_2)}{F(t_2)-F(t_1)} - \frac{\log[F(t_2)-F(t_1)]}{F(t_2)-F(t_1)}\bigg({\mathcal{M}}(t_1,t_2)+(\psi(t_2)-\psi(t_1))\frac{F(t_1)}{F(t_2)-F(t_1)}\bigg).
\end{array}\eeq
Here
\beqq\diy\mathcal{M}(t_1,t_2)=\bbE\Big[\psi(t_2)-\psi(X)|t_1\leq X \leq t_2\Big].\eeqq.
\end{rem}
We conclude the paper by using Theorem \ref{thm2.3.} for uniform RV, Theorem \ref{thm.sec2.2} in exponential form and WF $\varphi(x)=\diy\sum_{i=0}^n a_i x^i, \;a_i\in\bbR,\;\varphi(x)\geq 0$, recall also (\ref{Eq;1.7}, in order to explore some emerged inequalities.
\begin{cor}
{\rm (i)} For constant $0\leq a<b\leq 1$. Assume arbitrary function $f: \bbR\mapsto\bbR^+$ we get
\beqq \diy\int_a^b s\;f(s)\log\diy\frac{b-a}{s}\;\rd s\geq (b-a)\exp\diy\int_a^b \log \Big[s\;f(s)|\log \diy\frac{b-a}{s}|\Big]\;\frac{\rd s}{b-a}.\eeqq
{\rm (ii)} Consider constant $0\leq a<b\leq \infty$, $c,p\in\bbR^+$. Further set
\beqq\begin{array}{c}\overline{\gamma}_{p}(a,b)=\diy\int_{a}^{b}t^{p-1}e^{-t}\;\rd t=\diy\gamma(p,b)-\gamma(p,a),\;\; \hbox{by virtue of (\ref{def:gamma})}\\
\quad \diy\Delta_c(a,b)=e^{-a/c}-e^{-b/c}.\end{array}\eeqq
Then constants $\epsilon_0,\dots,\epsilon_n$, such that $\diy\sum_{i=0}^n \epsilon_i x^i\geq 0,\;x\in \bbR^+$ are satisfied in the inequality: 
\beq \label{Final:eq}\begin{array}{l} 2\diy\sum_{i=0}^n \big(c.\Delta_c(a,b)-\log \Delta_c(a,b)\big)c^i\;\epsilon_i\;\overline{\gamma}_{i+1}(a/c,b/c)-\sum_{i=0}^n \epsilon_i\; c^{i+1} 2^{-i}\overline\gamma_{i+1}(2a/c,2b/c)\\
\quad\quad\leq \diy\sum_{i=0}^n\big(2-\log \Delta_c(a,b)\big/(i+1)\big)\epsilon_i\;c^{i+1}\overline{\gamma}_{i+2}(a/c,b/c)\\
\qquad\qquad+\diy\log \Delta_c(a,b)\Big\{\sum_{i=0}^n \frac{\epsilon_i}{i+1}\big(a^{i+1}e^{-a/c}-b^{i+1}e^{-b/c}\big)\Big\}.\end{array}\eeq
Note that in special case $c=1$, $a\rightarrow 0$ and $b\rightarrow\infty$, final inequality (\ref{Final:eq}) takes the form
\beqq \sum_{i=0}^n \epsilon_i\;\Gamma(i+1)(1-2^{-i-1})\leq \sum_{i=0}^n \epsilon_i\;\Gamma(i+2). \eeqq
Here $\Gamma(.)=\overline{\gamma}_{.}(0,\infty)$ refers to Gamma function. 
\end{cor}
\vspace{0.5cm}
%\section{
{\emph{Acknowledgements --}}
 SYS thanks the CAPES PNPD-UFSCAR Foundation
for the financial support in the year 2014-5. SYS thanks
the Federal University of Sao Carlos, Department of Statistics, for hospitality during the year 2014-5.


\begin{thebibliography}{99}


\bibitem{CL} A. Di Crescenzo and M. Longobardi. On weighted residual and past entropies. {\it Scientiae Math. Japon}, \textbf{64} (2006), 255--266.

%\bibitem{E} N. Ebrahimi. How to measure uncertainty in the residual lifetime distribution. {\it Sankhya Series A}, \textbf{58} (1996), 48--56.


\bibitem{KRM} M. Khorashadizadeh, A. H. Rezaei Roknabadi and G. R. Mohtashami Borzadaran. Doubly truncated (interval) cumulative residual and past entropy. {\it Stat. and prob. Letters}, \textbf{83} (2013), 1464--1471.


\bibitem{M} F. Misagh, Y. Panahi, G. H. Yari and R. Shahi. Weighted cumulative entropy and its estimation. {\it Quality and Reliability (ICQR), IEEE International Conference on} (2011), 477--480.

\bibitem{MY} F. Misagh and G. H. Yari. On weighted interval entropy. {\it Statistics and Probability letters}, \textbf{81} (2011), 188--194.


\bibitem{NR} J. Navarro and J. M. Ruiz. Failure-rate function for doubly-truncated random variables. {\it IEEE Transactions on Reliability}, \textbf{45 (4)} (1996), 685--690.


\bibitem{RCVW} M. Rao, Y. Chen, B. C. Vemuri and F. Wang. Cumulative residual entropy, a new measure of information. {\it IEEE Transactions Information Theory}, \textbf{50} (2004), 1220--1228.


\bibitem{SSM} S. M. Sonuj, P. G. Sankaran, P.G. and S. S. Maya. Characterizations of life distributions using conditional expectations of doubly(interval) truncated random variables. {\it Communication Statistics- Theory Methods}, \textbf{38} (2009), 1441--1452.

\bibitem{SY} Y. Suhov and S. Yasaei Sekeh. Weighted cumulative entropies: An extension of CRE and CE. arXiv 1507.07051.

\bibitem{Ys} S. Yasaei Sekeh. A short note on estimation of WCRE and WCE. arXiv 1508.04742.

%\bibitem{Sh} C. E. Shannon. A mathematical theory of communication. {\it Bell System Technical J.}, \textbf{27} (1948), 379--423.

%\bibitem{BG} M. Belis and S. Guiasu.  A Quantitative and qualitative measure of information in cybernetic systems. {\it IEEE Trans. on Inf. Theory},  \textbf{14} (1968), 593--594.

%\bibitem{BNRA} F. Belzunce, J. Navarro, J. M. Ruiz and Y. Agulia. Some results on residual entropy function.
%{\it Metrika},  \textbf{59} (2004), 147--161.

%\bibitem{C} A. Clim. Weighted entropy with application. {\it Analele Universit\u{a}\c{t}ii Bucure\c{s}ti, Matematic\u{a}}, \textbf{Anul LVII} (2008), 223-231.

%\bibitem{CT1} T. Cover and J. Thomas. {\it Elements of Information Theory.} New York: Wiley, 2006.

%\bibitem{CT2} T.M. Cover and J.A. Thomas. Determinant inequalities via information theory. {\it SIAM J. Matrix
%Anal. and its Applicat.}, {\bf 9} (1988), 384--392.

%\bibitem{DCT} A. Dembo, T.M. Cover and J.A. Thomas. Information theoretic inequalities. {\it IEEE Trans.
%Inform. Theory}, \textbf{37} (1991), 1501--1518.

%\bibitem{DL} A. Di Crescenzo and M. Longobardi.  Entropy based measure of uncertainty in past lifetime distributions. {\it J. App. Prob.}, \textbf{39} (2002), no. 3, 434--440.

%\bibitem{DT} G. Dial and I. J. Taneja. t On weighted entropy of type ($\alpha$,$\beta$) and its generalizations.
%{\it Appl. Math.}, \textbf{26} (1981), 418--425.

%\bibitem{FS1} G. Frizelle and Y. M. Suhov. An entropic measurement of queueing behaviour in a class of manufacturing operations. {\it Proc. Royal Soc. A}, \textbf{457} (2001), 1579--1601

%\bibitem{FS2} G. Frizelle and Y. M. Suhov. The measurement of complexity in production and other
%commercial systems. {\it Proc. Royal Soc. A}, \textbf{464} (2008), 2649--2668.

%\bibitem{I} K. Ito. {\it Introduction to Probability Theory.} Cambridge: Cambridge University Press, 1984.

%\bibitem{G} S. Guiasu. Weighted entropy. {\it Report on Math. Physics}, \textbf{2} (1971), 165--179.

%\bibitem{JG} D. H. Johnson and R. M. Glantz.  When does interval coding occur? {\it Neurocopmuting},
%\textbf{59-60} (2004), 13--18.

%\bibitem{KSa} P. L. Kannappan and P. K. Sahoo. On the general solution of a functional equation
%connected to sum form information measures on open domain. {\it Math. Sci.}, \textbf{9} (1986), 545--550.

%\bibitem{K} J. N. Kapur. {\it Measures of Information and Their Applications.} Chapter 17,
%New Delhi: Wiley Eastern Limited, 1994.

%\bibitem{KS} M. Kelbert and Y. Suhov. {\it Information Theory and Coding by Example.} Cambridge: Cambridge University Press, 2013.

%\bibitem{MMN} K. Muandet, S. Marukatat and C. Nattee. Query selection via weighted entropy in graph-based semi-supervised classification. In:
%{\it Advances in Machine Learning.} Lecture Notes in Computer Science, \textbf{5828} (2009), pp. 278--292.

%\bibitem{NP} A. K. Nanda and P. Paul. Some results on generalized past entropy. {\it J. Statist. Plann. Inference}, \textbf{136} (2006), 3659--3674.

%\bibitem{PT} O. Parkash and H. C. Taneja. Characterization of the quantitative-qualitative measure of
%inaccuracy for discrete generalized probability distributions. {\it Commun. Statist. Theory Methods},
%\textbf{15} (1986), 3763--3771.

%\bibitem{ShMM} B. D. Sharma, J. Mitter and M. Mohan. On measure of  `useful` information. {\it Inform. Control} \textbf{39} (1978), 323--336.

%\bibitem{SiB} R. P. Singh and J. D. Bhardwaj. On parametric weighted information improvement. {\it Inf. Sci.} \textbf{59} (1992), 149--163.

%\bibitem{SrV} A. Sreevally and S. K. Varma. Generating measure of cross entropy by using measure of weighted entropy. {\it Soochow Journal of Mathematics}, \textbf{30} (2004), no. 2, 237--243.

%\bibitem{S} A. Srivastava. Some new bounds of weighted entropy measures. {\it Cybernetics and Information Technologies}, \textbf{11} (2011), no. 3, 60--65.

%\bibitem{TCJ} R. K. Tuteja, Sh. Chaudhary and P. Jain. Weighted entropy of orders $\alpha$ and type $\beta$ information energy. {\it Soochow Journal
%of Mathematic} \textbf{19} (1993), no. 2, 129-138.

\end{thebibliography}
\end{document}